\documentclass[conference]{IEEEtran}
\IEEEoverridecommandlockouts
\usepackage{cite}
\usepackage{amsmath,amssymb,amsfonts}
\usepackage{algorithmic}
\usepackage{graphicx}
\usepackage{textcomp}
\usepackage{xcolor}
\def\BibTeX{{\rm B\kern-.05em{\sc i\kern-.025em b}\kern-.08em
    T\kern-.1667em\lower.7ex\hbox{E}\kern-.125emX}}
\begin{document}

\title{Analyzing Lack of Concordance Between the Proteome and Transcriptome in Paired scRNA-Seq and Multiplexed Spatial Proteomics}

\author{\IEEEauthorblockN{Jai Prakash Veerla\IEEEauthorrefmark{1}\IEEEauthorrefmark{2}, Jillur Rahman Saurav\IEEEauthorrefmark{1}\IEEEauthorrefmark{2}, Michael Robben\IEEEauthorrefmark{1}\IEEEauthorrefmark{2}, Jacob M. Luber\IEEEauthorrefmark{1}\IEEEauthorrefmark{2}}
\IEEEauthorblockA{\IEEEauthorrefmark{1}Department of Computer Science, University of Texas at Arlington}
\IEEEauthorblockA{\IEEEauthorrefmark{2} Multi-Interprofessional Center for Health Informatics, University of Texas at Arlington }

\IEEEauthorblockA{Email: {jxv6663@mavs.uta.edu,\{mdjillurrahman.saurav, michael.robben, jacob.luber}\}@uta.edu
}
}

\maketitle

\begin{abstract}
In this study, we analyze discordance between the transcriptome and proteome using paired scRNA-Seq and multiplexed spatial proteomics data from HuBMAP. Our findings highlight persistent transcripts in key immune markers, including CD45-RO, Ki67, CD45, CD20, and HLA-DR. CD45-RO is consistently expressed in memory T cells, while Ki67, associated with cell proliferation, also displays sustained expression. Furthermore, HLA-DR, part of the MHC class II molecules, demonstrates continuous expression, possibly crucial for APCs to trigger an effective immune response. This investigation provides novel insights into the complexity of gene expression regulation and protein function.

\end{abstract}


\section{Introduction}

Pioneering advancements in multi-omics technologies have revolutionized our capacity to scrutinize gene expression at the cellular level \cite{goltsev2018deep,robben2023state}. Through the integration of scRNA-Seq and multiplexed imaging like CO-Detection by indEXing (CODEX), we are now able to track both transcripts and corresponding protein expressions. However, inconsistencies have been observed between spatial proteomics and scRNA-Seq data, potentially arising due to batch or technical effects during data collection. 

Despite considerable progress in scRNA-Seq technologies, data generated often contains substantial noise, with significant dropout events leading to undetected transcripts due to either biological or technical issues. This typically results in an overrepresentation of zero values, necessitating the use of zero-inflation statistical models \cite{jiang2022statistics}. Two primary sources of these zero values exist: technical zeros, when a gene is expressed but remains undetected, and biological zeros, occurring when a gene is simply not expressed in a cell.

\begin{figure} [ht!]
  \centering
  \includegraphics[width=\linewidth]{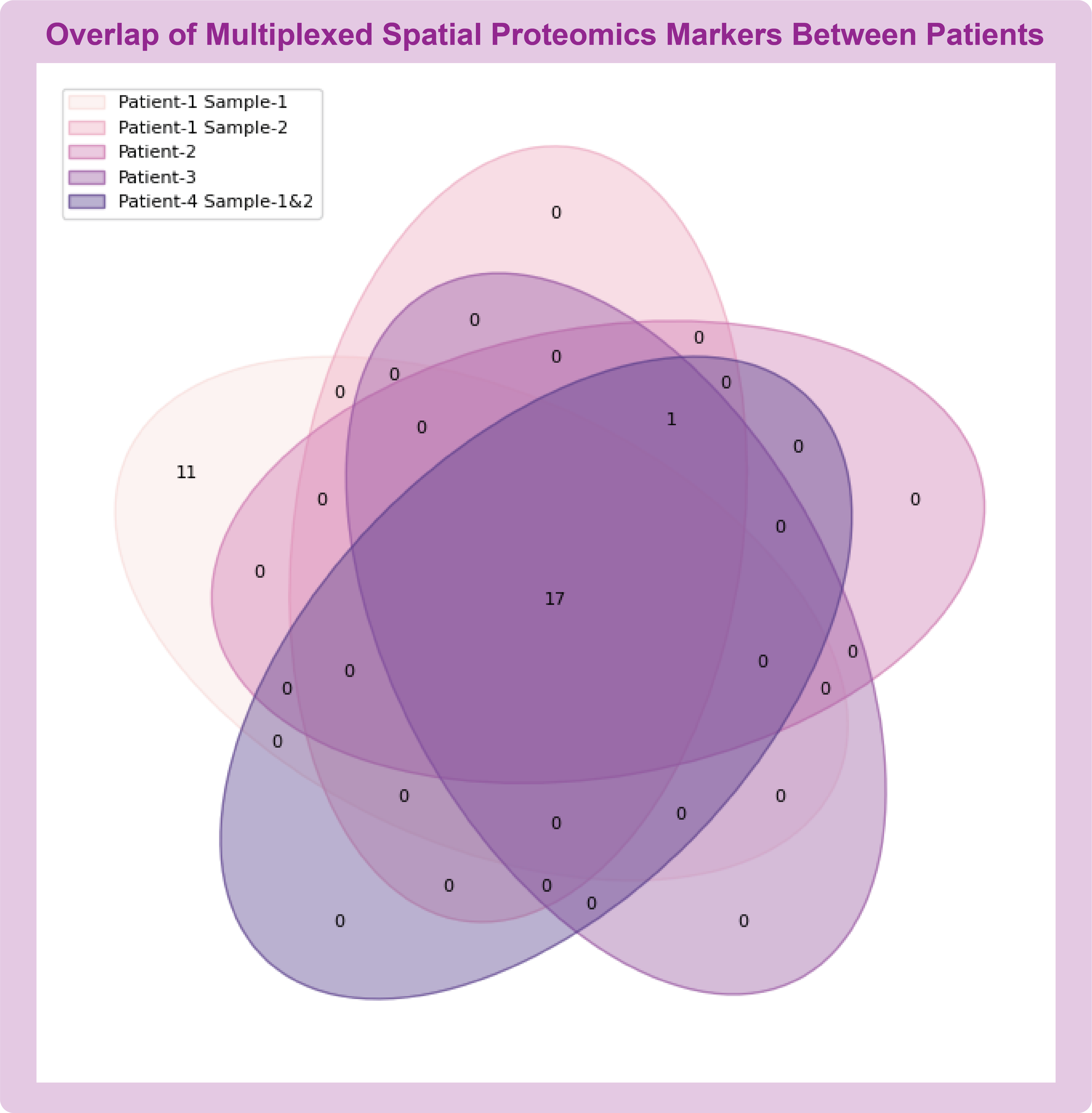}
  \caption{Overlap of Multiplexed Spatial Proteomics Markers Between Patients.}
  \label{fig:overlap}
\end{figure}

By visualizing and measuring the activity of various gene expressions in CODEX images, compared with scRNA-Seq count matrix data from HuBMAP, we aim to explore the role of batch and technical effects on transcriptional bursting. Our preliminary findings indicate certain genes showing no expression in scRNA-Seq data, yet present in CODEX images, and others where transcripts are overexpressed relative to proteins. This hints at the potential to measure transcriptional bursting across platforms such as scRNA-Seq and spatial proteomics, using paired scRNA-Seq SALMON data and CODEX imaging data from HuBMAP across multiple patient samples.

\section{Methods}

\begin{figure} [ht!]
  \centering
  \includegraphics[width=\linewidth]{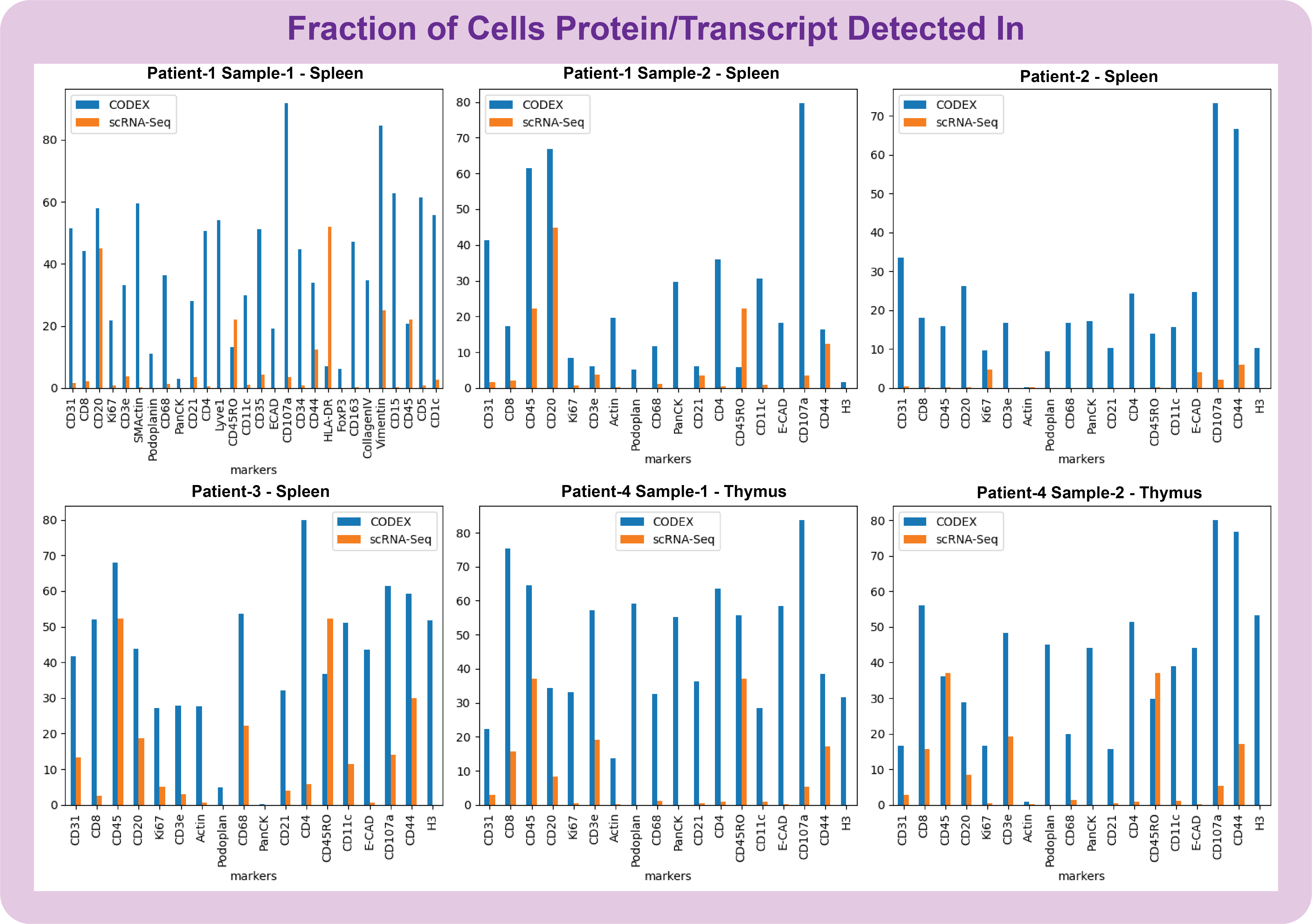}
  \caption{Fraction of Cells that Proteins and Transcripts of Interest are Detected.}
  \label{fig:no_concordance}
\end{figure}
We began our investigation by obtaining both transcriptomic and proteomic data from HuBMAP, specifically, paired scRNA-Seq (Salmon) \cite{patro2015salmon} and CODEX datasets \cite{dayao2022membrane} performed on tissue from the same donors. This resulted in the acquisition of four paired datasets collected by the University of Florida from organs including the spleen and thymus.

The scRNA-Seq (Salmon) dataset was processed using the "vpolo" tool to retrieve the gene expression matrix, which is characterized by rows of cells and columns of genes, with values representing the gene expression of each cell. Given the zero-inflated nature of scRNA-Seq data, we noted a significant number of zeros in the matrix.

For proteomic data analysis, we selected a subset of protein markers used in CODEX corresponding to the genes in our gene expression matrix. We utilized CytoKit \cite{czech2019cytokit} for image segmentation to identify protein marker expression. Following the acquisition of both the gene expression matrix from scRNA-Seq and protein marker expression from CODEX, we conducted comparative visualizations. All code is available at: https://github.com/jacobluber/TranscriptionalBursting.
% This resulted in the collection of 13 donor datasets from HuBMAP created by the University of Florida and Stanford University for organs such as the spleen, thymus, small intestine, and large intestine. 

\section{Results}
\begin{figure} [ht!]
  \centering
  \includegraphics[width=\linewidth]{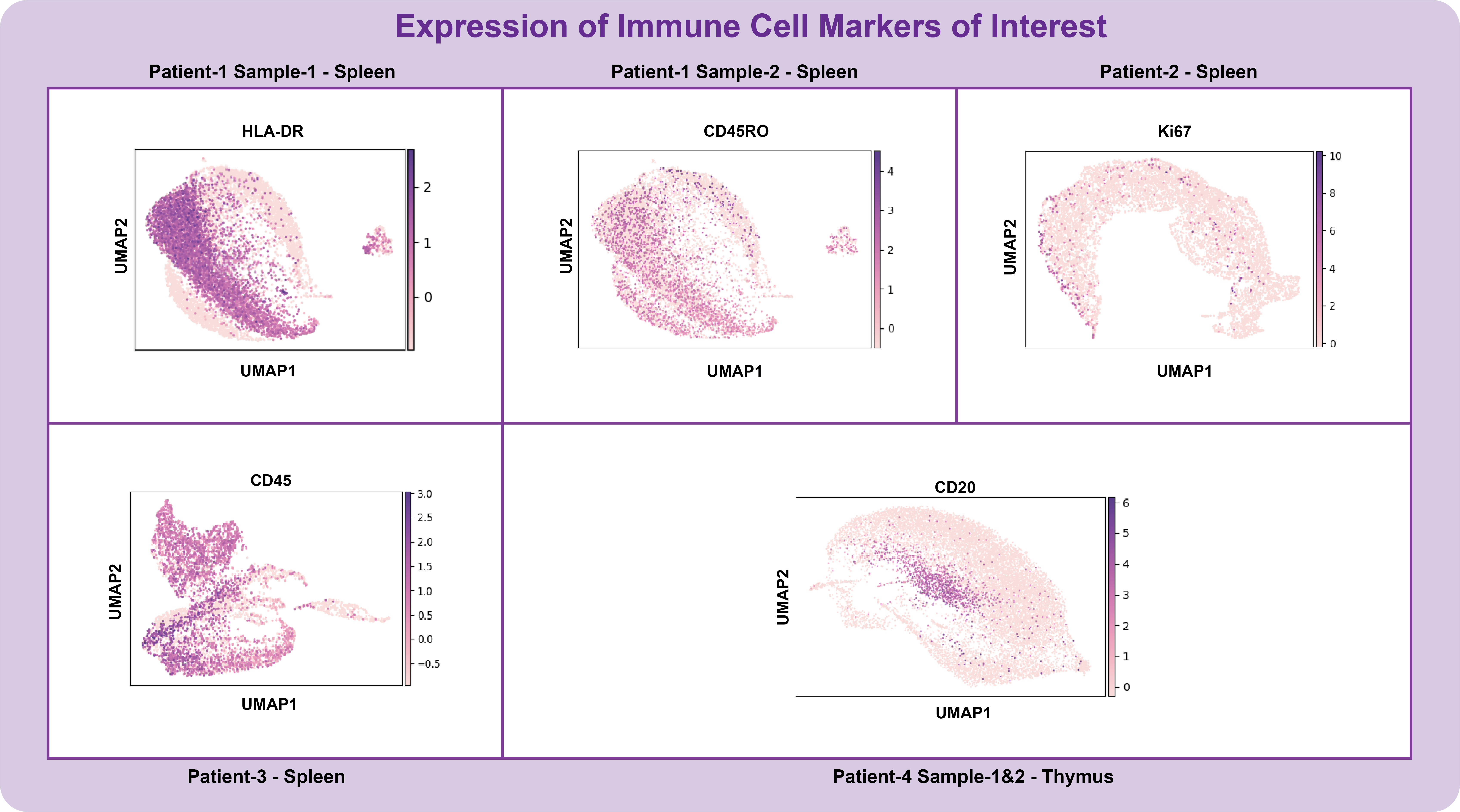}
  \caption{Expression of Immune Cell Markers of Interest.}
  \label{fig:of_interest}
\end{figure}

Several immune cell surface markers, such as CD45-RO, Ki67, CD45, and CD20, showed persistent levels of transcript expression (Fig. \ref{fig:no_concordance}). These markers play crucial roles in immune function and may require persistent transcripts to fulfill their functions effectively (Fig. \ref{fig:of_interest}). For example, CD45-RO, an isoform of CD45, is typically expressed on memory T cells. Memory T cells, which have previously encountered an antigen, can respond more quickly upon subsequent encounters with the same antigen. The persistent expression of CD45-RO is necessary to maintain their memory function.

Ki67, another marker that displayed sustained expression (Fig. \ref{fig:no_concordance}), is a nuclear protein associated with cellular proliferation. In immune cells, especially lymphocytes, Ki67 is used as a marker for cell division and growth. Continuous transcript expression of Ki67 ensures that rapidly dividing cells can maintain their proliferation.

We also observed sustained expression of HLA-DR (Fig. \ref{fig:no_concordance}, Fig. \ref{fig:of_interest}), which is part of the Major Histocompatibility Complex (MHC) class II molecules. APCs present antigens to CD4+ T cells, initiating an immune response. To carry out this function effectively, these cells may need to constantly express HLA-DR on their surface, requiring stable transcripts for HLA-DR.

The concept of MHC complex segregation in the context of immune responses involves the presentation of MHC molecules loaded with foreign peptides on the cell surface to be recognized by T cells. The segregation of MHC complexes may increase the diversity of antigens presented, thereby enhancing the likelihood of triggering an appropriate immune response. Therefore, the presence of persistent transcripts for MHC molecules enables immune cells to respond quickly to pathogens by ensuring the continuous availability of proteins required for immune responses.

\section{Conclusion}
As a future direction, combining multiplexed imaging techniques such as CODEX with single-cell RNA sequencing (scRNA-Seq) could provide a more detailed understanding of immune cell dynamics. Pseudotime algorithms, which infer dynamic changes of single cells along a hypothetical timeline, would allow us to track temporal changes in gene expression and visualize transcriptional bursting events. This approach could reveal how immune response is regulated at the single-cell level, highlighting key temporal changes in immune cell markers like CD45-RO, Ki67, CD45, CD20, and HLA-DR. Consequently, these integrated techniques have the potential to revolutionize our understanding of immune system regulation and provide insights for therapeutic interventions.

\section*{Acknowledgment}

 This work was supported by a University of Texas System Rising STARs Award (J.M.L) and the CPRIT First Time Faculty Award (J.M.L)

% The 
% sentence punctuation follows the bracket \cite{b2}. Refer simply to the reference 
% number, as in \cite{b3}---do not use ``Ref. \cite{b3}'' or ``reference \cite{b3}'' except at 
% the beginning of a sentence: ``Reference \cite{b3} was the first $\ldots$''

% Number footnotes separately in superscripts. Place the actual footnote at 
% the bottom of the column in which it was cited. Do not put footnotes in the 
% abstract or reference list. Use letters for table footnotes.

% Unless there are six authors or more give all authors' names; do not use 
% ``et al.''. Papers that have not been published, even if they have been 
% submitted for publication, should be cited as ``unpublished'' \cite{b4}. Papers 
% that have been accepted for publication should be cited as ``in press'' \cite{b5}. 
% Capitalize only the first word in a paper title, except for proper nouns and 
% element symbols.

% For papers published in translation journals, please give the English 
% citation first, followed by the original foreign-language citation \cite{b6}.

\bibliographystyle{ieeetr}
\bibliography{conference_101719.bib}

\begin{thebibliography}{1}

\bibitem{goltsev2018deep}
Y.~Goltsev, N.~Samusik, J.~Kennedy-Darling, S.~Bhate, M.~Hale, G.~Vazquez,
  S.~Black, and G.~P. Nolan, ``Deep profiling of mouse splenic architecture
  with codex multiplexed imaging,'' {\em Cell}, vol.~174, no.~4, pp.~968--981,
  2018.

\bibitem{robben2023state}
M.~Robben, A.~Hajighasemi, M.~S. Nasr, J.~P. Veerla, A.~M. Alsup, B.~Rout,
  H.~H. Shang, K.~Fowlds, P.~B. Malidarreh, P.~Koomey, M.~J.~R. Saurav, and
  J.~M. Luber, ``The state of applying artificial intelligence to tissue
  imaging for cancer research and early detection,'' 2023.

\bibitem{jiang2022statistics}
R.~Jiang, T.~Sun, D.~Song, and J.~J. Li, ``Statistics or biology: the
  zero-inflation controversy about scrna-seq data,'' {\em Genome biology},
  vol.~23, no.~1, pp.~1--24, 2022.

\bibitem{patro2015salmon}
R.~Patro, G.~Duggal, and C.~Kingsford, ``Salmon: accurate, versatile and
  ultrafast quantification from rna-seq data using lightweight-alignment,''
  {\em BioRxiv}, vol.~10, p.~021592, 2015.

\bibitem{dayao2022membrane}
M.~T. Dayao, M.~Brusko, C.~Wasserfall, and Z.~Bar-Joseph, ``Membrane marker
  selection for segmenting single cell spatial proteomics data,'' {\em Nature
  communications}, vol.~13, no.~1, p.~1999, 2022.

\bibitem{czech2019cytokit}
E.~Czech, B.~A. Aksoy, P.~Aksoy, and J.~Hammerbacher, ``Cytokit: a single-cell
  analysis toolkit for high dimensional fluorescent microscopy imaging,'' {\em
  BMC bioinformatics}, vol.~20, no.~1, pp.~1--13, 2019.

\end{thebibliography}

\end{document}